\documentclass[aps,prl,reprint]{revtex4-2}

\usepackage{graphicx}
\usepackage{dcolumn}
\usepackage{bm}
\usepackage{physics}

\begin{document}

\preprint{}

\title{Superscattering Empowered by Bound States in the Continuum}

\author{Adri\`a Can\'os Valero,$^1$ Hadi~K.~Shamkhi,$^1$ Anton S. Kupriianov,$^2$ Vladimir R. Tuz,$^3$ Alexander A. Pavlov,$^4$ Dmitrii Redka,$^5$ Vjaceslavs Bobrovs,$^6$ Yuri S. Kivshar,$^7$ and Alexander~S.~Shalin$^{1,6,8}$}

\affiliation{$^1$ITMO University, St.~Petersburg 197101, Russia}
\affiliation{$^2$College of Physics, Jilin University, Changchun 130012, China} 
\affiliation{$^3$State Key Laboratory of Integrated Optoelectronics, College of Electronic Science and Engineering, International Center of Future Science, Jilin University, Changchun 130012, China}
\affiliation{$^4$Institute of Nanotechnology of Microelectronics of the Russian Academy of Sciences, Moscow 119991, Russia}
\affiliation{$^5$Electrotechnical University LETI, St. Petersburg 197376, Russia}
\affiliation{$^6$Riga Technical University, Institute of Telecommunications, Riga 1048, Latvia}
\affiliation{$^7$Nonlinear Physics Centre, Australian National University, Canberra ACT 2601, Australia}
\affiliation{$^8$Kotelnikov Institute of Radio Engineering and Electronics of Russian Academy of Sciences (Ulyanovsk branch), Ulyanovsk 432000, Russia}

\begin{abstract}
We uncover a novel mechanism for superscattering of subwavelength resonators closely associated with the physics of bound states in the continuum.  We demonstrate that the enhanced scattering occurs as a consequence of constructive interference within the Friedrich-Wintgen mechanism of interfering resonances, and it may exceed the currently established limit for the cross-section of a single open scattering channel.  We develop a non-Hermitian model to describe interfering resonances of quasi-normal modes to show that this effect can only occur for scatterers violating the spherical symmetry, and therefore it cannot be predicted with the classical Mie solutions. Our results reveal unusual physics of non-Hermitian systems having important implications for functional metadevices.  
\end{abstract}

\maketitle

{\em Introduction.} Non-Hermitian physics offers a wide range of unusual phenomena not accessible for purely Hermitian systems \cite{el2019dawn}. In recent years, there has been a tremendous progress in implementations of non-Hermitian platforms in optics
leading to the discovery of many interesting effects naturally occurring in lossy or gain-compensated optical structures. Being motivated by the studies of parity-time ($\mathcal{PT}$) symmetric systems, the novel field of non-Hermitian photonics emerged~\cite{el2019dawn}. The latter takes advantage of new degrees of freedom offered by complex energy landscapes~\cite{hodaei2017enhanced,feng2014single,feng2013experimental,koshelev2020subwavelength,chebykin2015strong}. The advancements are particularly exciting for nanophotonics, since new concepts enable the study of unconventional regimes of light-matter interaction such as exceptional points \cite{huang2017unidirectional,park2020symmetry} or dark states \cite{valero2020theory,limonov2017fano,gongora2017anapole}. 

Unlike many $\mathcal{PT}$-symmetric systems, the eigenvalues of an isolated optical nanoresonator with uncompensated radiative losses are complex. Nevertheless, they can be controlled by engineering cavity parameters achieving quasi-BIC modes with ultrahigh $\mathcal{Q}$-factors and strong localization \cite{rybin2017high}. This regime arises due to destructive interference within the modes of the same radiation channel, owing to the Friedrich-Wintgen (FW)  mechanism of interfering resonances~\cite{friedrich1985}. While a bimodal system coupled to one channel of the continuum is well understood, the subtleties underlying multiple channel interactions are yet to be exploited in photonics. $\emph{Destructive}$ interference leads to a quasi-BIC regime and the suppression of radiation in one channel. Here, we pose the question whether $\emph{constructive}$ interference of a quasi-BIC state and a low-$\mathcal{Q}$ mode in a multi-channel structure can boost radiation beyond the limit for an isotropic scatterer, achieving the superscattering regime \cite{qian2019experimental, Tuz_PhysRevApplied_2020}.  Up to now, superscattering was known to originate only from an accidental degeneracy of the modes~\cite{ruan2010superscattering}
exceeding the single-channel cross-section limited to $\sigma_{0}=0.5\lambda^2(2\ell+1)/\pi$. 

\begin{figure}[!t]
    \centering
    \includegraphics[scale=0.2]{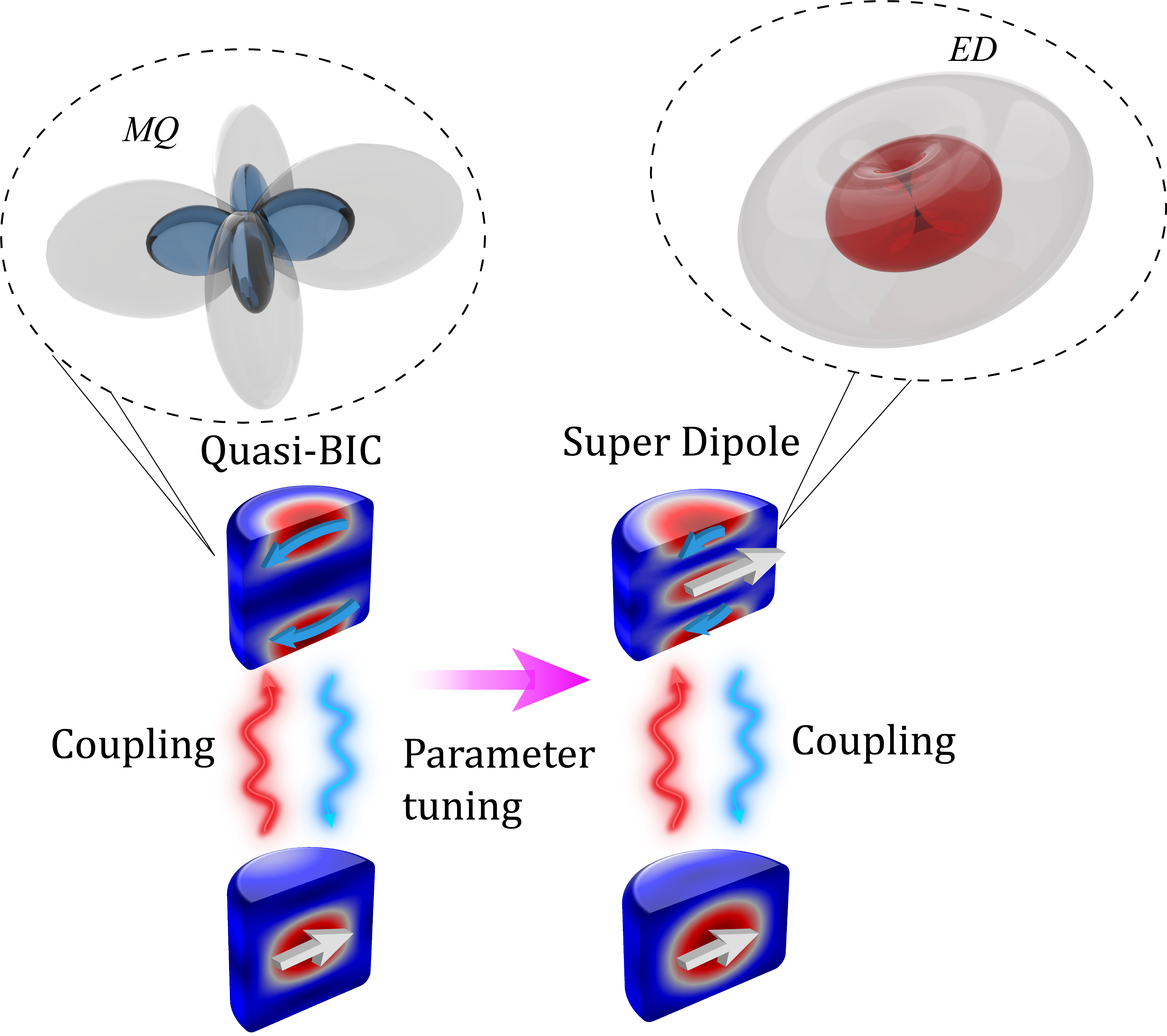}
    \caption{Schematic of a two-mode system leading to BIC-driven superscattering. Two coupled quasi-normal modes drastically reshape their near fields and scattering patterns as a function of a single tuning parameter. The blue and red scattering patterns indicate the previously known bounds for dipole scattering from a subwavelength nanoparticle~\cite{ruan2010superscattering,FanTransSuperscattering}. The interfering resonances induce the energy redistribution between multipolar channels creating a super dipole mode, and thus almost doubling the dipole scattering limit.}
    \label{fig_1}
\end{figure}

In this Letter, we demonstrate another important feature of BICs in open resonators. More specifically, we reveal that interfering resonances can lead to previously unknown regimes of superscattering, once the spherical symmetry is broken, as depicted in Fig.~\ref{fig_1}. First, we introduce a simple model to describe this effect analytically, and then employ cavity perturbation theory for quasi-normal modes (QNMs) of non-Hermitian resonators \cite{LalanneReview,yan2020shape}.  We reveal that mode coupling induces energy redistribution within two radiation channels allowing to overcome the single-channel scattering limit \emph{within the same channel}, and/or control the quality factor ($\mathcal{Q}$-factor) along with the multipolar content of the resonances.

We start from a two-level non-Hermitian Hamiltonian:
\begin{equation} \label{eq:1}
    \mathcal{H}_0=\left(\begin{array}{cc}
         \omega_1-i\gamma_1&g  \\
         g&\omega_2-i\gamma_2 
    \end{array}\right),
\end{equation}
where the two modes with complex eigenfrequencies $\tilde{\omega}_i=\omega_i-i\gamma_i$ are coupled through $g=\kappa-i\nu$. The eigenmodes of $\mathcal{H}_0$ decay from the resonator after some finite time dictated by $\gamma_i$ and $g$. Due to $g$, the original states $\ket{1,2}$ hybridize into the dressed states $\ket{u,d}$, with higher and lower energy, respectively. $\mathcal{H}_0$ interacts with a continuum of open channels which determine the form of the $\gamma_i$ and $\nu$. In the case of an isolated resonator interacting with a plane wave, those channels are usually associated with a basis of vector spherical multipoles (electric and magnetic dipoles, quadrupoles, etc.). For initial considerations, we can assume that the resonant modes can be energy-normalized in order to make use of an approximate, but highly insightful set of equations known as temporal coupled mode theory (TCMT):

\begin{equation}
    \label{eq:5}
    \frac{d\mathbf{\ket{\Psi}}}{dt}=-iH\mathbf{\ket{\Psi}}+i\sqrt{2}D^T\mathbf{s}^+,
\end{equation}
where $\mathbf{\ket{\Psi}}=(\ket{u},\ket{d})^T$, are the eigenvectors, normalized by the internal electromagnetic energy, and $D$ is a matrix coupling the incoming channels $\mathbf{s}^+=(s^+_1,s_2^+)^T$ to $\ket{\mathbf{\Psi}}$. The $D_{ij} $ component relates the eigenmode $i=u,d$, to the channel $j=1,2$. Reciprocity and energy conservation requirements lead to a relation between the components of $D$ and $\gamma_i,\nu$ (see Sec. S1 in the Supplemental Material \cite{Suppl_Mat}). The TCMT describes approximately the dynamics of the QNMs characterizing the stationary response of the photonic system. They are strictly valid under the assumptions of weak coupling $(g\rightarrow 0)$ and highly confined resonances (small non-Hermitian perturbations) \cite{haus1984waves}. They have ample use in the photonics community \cite{fan2003temporal, volkovskaya2020multipolar, Tuz_PhysRevApplied_2019, gorkunov2020metasurfaces}, and have been shown to describe well the formation of quasi-BICs \cite{volkovskaya2020multipolar}. Here we use it only for illustration purposes, and verify our results with a rigorous cavity perturbation theory in a realistic nanoresonator.

The modes transfer the energy from the incoming to the outgoing channels according to $\mathbf{s}^{-}=\mathbf{s}^++D\ket{\mathbf{\Psi}}$. Under harmonic time dependence analytical expressions for the transmission matrix $\mathcal{T(\omega)}$, relating the incident and scattered field amplitudes (see Sec. S1 in the Supplemental Material \cite{Suppl_Mat}) can be derived. The contributions from each channel to the scattering cross-section are then readily obtained as $\sigma_i(\omega)=2\sum_{j}\mathcal{T}_{ij}(\omega)s^+_j$, where the factor of 2 stems from the incoming and outgoing components contained in a plane wave \cite{ruan2010superscattering}.

\begin{figure*}[!htb]
\centering
    \includegraphics[width=1.0\textwidth]{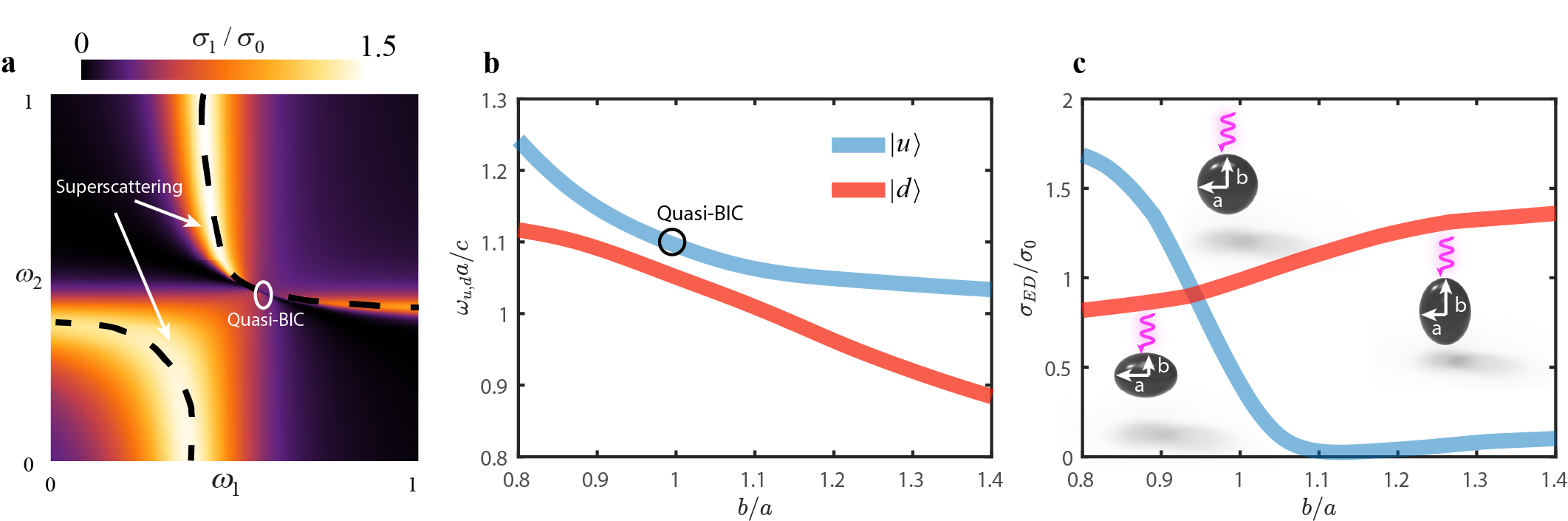}
    \caption{(a) Analytical model of the normalized scattering contribution of the first channel calculated with TCMT, featuring a quasi-BIC mode at $(\omega_1,\omega_2)=(0.6,0.5)$. The black dashed lines track the path followed by the real parts of the eigenfrequencies of the bimodal system. The single channel limit is observed to be exceeded in both high and low energy branches. (b) Evolution of the eigenfrequencies of the ED and MQ modes of a silicon nanosphere once the rotational symmetry along z is broken, rendering it an ellipsoid. (c) Calculated scattering cross-section of the ED channel at the peaks of the two resonances as a function of ellipticity, under normally incident, linearly polarized plane wave illumination. For $b/a=1$ (a sphere), the scattering cross-section is bounded to 1, and the upper branch features a quasi-BIC. When the ellipsoid is oblate (left inset), the upper branch becomes a super-dipole mode, and vice versa for the lower branch (right inset).  }
    \label{fig_2}
\end{figure*}

{\em Single channel superscattering.} If the off-diagonal components of the $D_{ij}$ are nonzero, the $\mathcal{T}$-matrix is non-diagonal and the modes are allowed to couple to at least two different channels. This must be the case for resonators without spherical symmetry \cite{krasikov2021,mishchenko1996t}. As a consequence, channel mixing is allowed within the scatterer, and the scattering cross-section of an object can take in principle arbitrarily large values \cite{ruan2012temporal}. In accordance with the FW mechanism, introducing coupling between modes induces both destructive and constructive interference that can be suitably designed to control both the $\mathcal{Q}$-factors and the multipolar characteristics of the resonances by varying a limited set of system parameters. We are interested in the possibility to harness the latter to enhance scattering \emph{in one channel} beyond the conventionally accepted limit. 

As an exemplary case, we further reduce the model by assuming $\gamma_1\approx\nu\approx{}d_1^2$, and $\gamma_2=d_1^2+d_0^2$, where $d_1,d_0$ are the direct coupling terms to channels 1 and 2, respectively. With these assumptions, we impose that the two modes leak to channel 1, but only one mode leaks to channel 2. Therefore, a quasi-BIC can only potentially appear in channel 1 \cite{remacle1990}. We study the behavior of the eigenfrequencies, and the evolution of the scattering cross-section at the resonances for channel 1 [Fig. \ref{fig_2}(a)]. When $\omega_1=\omega_2$, the imaginary part of the eigenfrequencies takes the simple form
\begin{equation} \label{eq:3}
    2\Im(\bar{\omega}_{u,d})\approx-d_1^2\Big[(2+\alpha^2)\mp{}\sqrt{4+\alpha^4}\Big]+O(\kappa^2),
\end{equation}
where $\alpha=d_0/d_1$. The upper branch then displays a considerably reduced imaginary part, implying a very high finesse of the resonance lineshape, and very large $\mathcal{Q}$-factor. This results in a quasi-BIC, effectively becoming `dark' to the incoming radiation in channel 1. The same mechanism implies a considerable brightening of the lower energy branch, while the reduction of the $\mathcal{Q}$-factor does not necessarily imply a reduction of the scattering cross-section; at $\omega_1$ the latter is given by
\begin{align}
\label{eq:4}
    \frac{\sigma_1}{\sigma_0}=\frac{4\gamma_1^2(1+\alpha)}{4\gamma_1^2+\kappa^2}+O(\alpha^2),
\end{align}
(see proof in Sec. S2 in the Supplemental Material \cite{Suppl_Mat}). If no channel mixing takes place, $\alpha=0$, following immediately that $\sigma_d\leq\sigma_0$. Once inter-channel mixing is enabled, however, $\sigma_d$ can exceed $\sigma_0$, overcoming the single channel limit \emph{within the same channel}. Similarly, it can be appreciated in Fig. \ref{fig_2}(a) that constructive interference between the modes allows superscattering to occur in both high and low energy branches, enabling the formation of `super-multipoles'. In the vicinity of the quasi-BIC, the destructive interference prevents the system from radiating in one of the channels and enhances scattering on the other one, with inherently larger $\mathcal{Q}$-factor.  

This simple analysis already shows interesting possibilities that could be realized, such as broadband superscattering making use of the low energy branch, or enhanced light-matter interactions by exploiting the higher $\mathcal{Q}$-factor of the upper one. Alternatively, the parameters of the Hamiltonian can be adjusted in such a way as to leak energy primarily towards one specific multipolar channel, enabling the radiation of `super-multipoles' with well-defined parity and angular momentum from a subwavelength antenna.

The fact that the limit can be overcome within the channel itself has remained unexplored until now. In the following, we will show that this new type of superscattering can be traced back to the hybridization of two Mie modes, whose near and far field coupling induces nonzero off-diagonal elements of the $\mathcal{T}$-matrix. Irregardless of the system, if only two channels with incoming amplitudes $a_1,a_2$ are coupled, $\sigma_1$ is written as $\sigma_1=\norm{a_1\mathcal{T}_{11}+a_2\mathcal{T}_{12}}^2$. The sum of the contributions is maximized when the diagonal and off-diagonal terms interfere \emph{constructively}, i.e. when the two contributing terms are in phase (we provide a more robust criteria in Sec. S3 in the Supplemental Material \cite{Suppl_Mat}). The generalization to arbitrary channels follows immediately from the previous.

{\em Subwavelength nanoresonators.}
First, we consider a Si nanosphere embedded in air, with radius $a=100$~nm. The sphere supports two resonances (see Supplemental Information S4 \cite{Suppl_Mat}) radiating as the electric dipole (ED, $\ket{d}$) and the magnetic quadrupole (MQ, $\ket{u}$). Due to the spherical symmetry, the two resonances radiating in different channels cannot interact with each other. Thus, the scattering of the ED channel is limited to $\sigma_0$. In the sphere, $\ket{u}$ effectively corresponds to a quasi-BIC, since it is completely decoupled from the ED channel. We now perform a controlled symmetry breaking by changing the ratio between the two orthogonal axis of the resonator [refer to schematic insets in Fig. \ref{fig_2}(c)]. In this scenario, the symmetry is reduced from the point group $C_{\infty{}h}$ to $D_{\infty{}h}$, allowing modes with even (or odd) parity to interact \cite{xiong2020constraints}. The result is the appearance of an avoided crossing in the eigenfrequency spectrum [Fig. \ref{fig_2}(b)], and a leakage of energy from the quasi-BIC mode to the ED channel. When the spheroid is oblate, the ED channel at $\omega_u$ almost doubles its allowed bound [Fig. \ref{fig_2}(c)], while the lower branch behaves similarly when the spheroid becomes prolate.
\begin{figure}[!htb]
    \centering
    \includegraphics[width=0.4\textwidth]{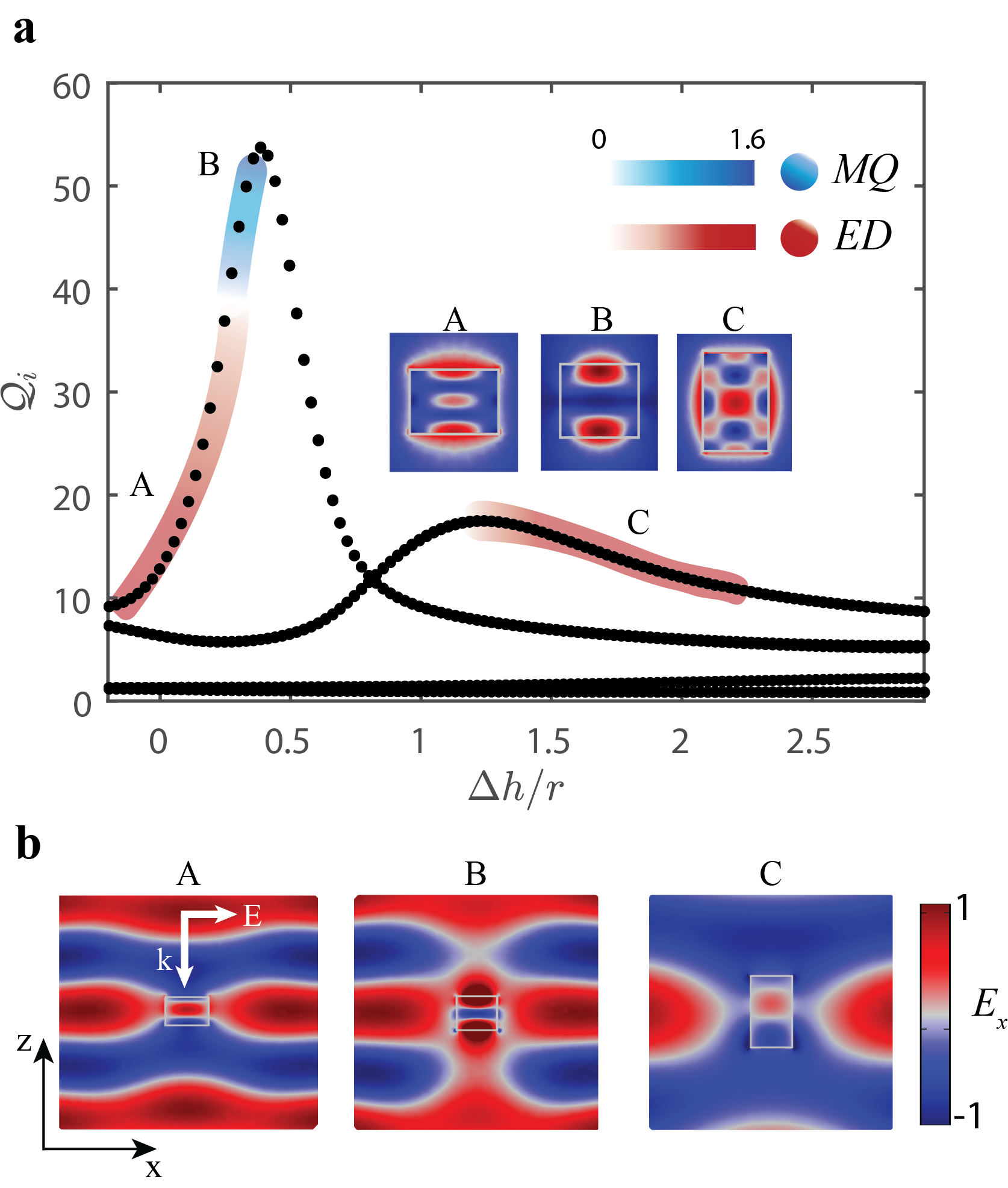}
    \caption{(a) Calculated $\mathcal{Q}$-factors of the involved QNMs, analytically obtained via exact first order cavity perturbation theory. The colored regions indicate where superscattering is achieved within a single channel for the electric dipole $ED$ (red) or for the magnetic quadrupole $MQ$ (blue). All the values are normalized by the dipole channel limit \cite{FanTransSuperscattering}. Insets show the electric field norms in the $x$-$z$ plane of the QNMs at points A, B, C. (b) Shows the $E_x$ component of the total field at the superscattering points A-C indicated in (a), under $x$-polarized plane wave illumination.}
    \label{fig_3}
\end{figure}

\begin{figure*}[!htb]
    \centering
    \includegraphics[width=1.0\textwidth]{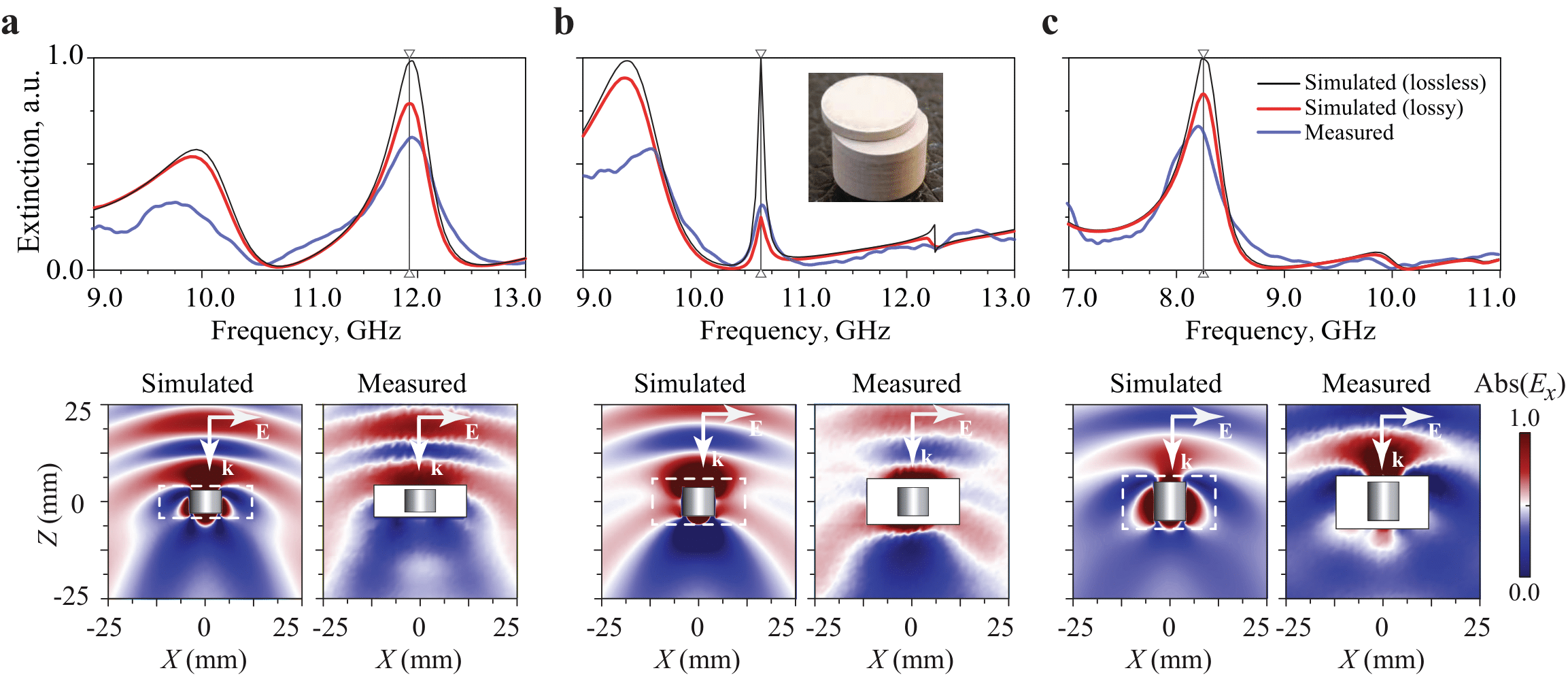}
    \caption{Comparison of the simulated and measured total extinction cross-section and scattered electric near-field patterns. Insets show an example of the experimental particle and the electric field norms in the $x$-$z$ plane of the QNMs at points A, B, and C. In the near-field patterns, the white spots around the disks correspond to the physically inaccessible zones for the measurement. The disks aspect ratios are: (a) $\Delta h/r=0.25$, (b) $\Delta h/r=0.625$, and (c) $\Delta h/r=1.25$.}
    \label{fig_5}
\end{figure*}

This example illustrates nicely the formation of a super ED from a symmetry-breaking perturbation. However, spheroids are in general not fabrication-friendly at the nanoscale. Instead, it is possible to design a super ED by making use of the QNMs excited in a silicon nanorod under normally incident plane wave illumination, since they also belong to the symmetry group $D_{\infty{}h}$. To do so, we perform an axial deformation parameterized by $\Delta{}h$, starting from a chosen height of $h_0=180~\text{nm}$, for which two modes radiating as ED and MQ are spectrally close. To describe the coupling of the QNMs in parameter space, the eigenvalues of the perturbed Hamiltonian are obtained analytically from first order cavity perturbation theory by solving a generalized eigenvalue problem due to a perturbation in the original eigenfrequencies given by \cite{yan2020shape}
\begin{equation} \label{eq:6}
    \mathcal{H}^{ij}_p=\Delta{}h\delta\varepsilon\braket{\mathbf{(\tilde{E}}_i^\textrm{out})^*}{{\mathbf{\tilde{E}}_j^\textrm{in}}}.
\end{equation}
where the Bra-Ket notation denotes the usual inner product over the resonator volume, and $i,j$ index the QNMs. From an initially large numerical set of QNMs, we isolate those that play a role in the coupling process according to the criteria ${|\Delta\hat{\omega}|}/{|\hat{\omega}_0|}\gg|\mathcal{H}_{p,0i}|$. From that reduced set, we discard the ones that are prevented by symmetry from radiating in the ED channel. As a result we obtain a system of two coupled resonant QNMs of relatively high $\mathcal{Q}$-factor which can also interact with two additional QNMs of very low $\mathcal{Q}$, associated with the scattering background. While the role of the background modes is usually disregarded in most analysis since their spectral signature is barely appreciable in the observables, the correct eigenfrequencies and the scattering response of the system cannot be well predicted without taking them into account (the field distributions and contributions to extinction of all the modes involved are shown in Sec. S5 of the Supplemental Information \cite{Suppl_Mat}).

In Fig.~\ref{fig_3}(a), the obtained $\mathcal{Q}$-factors of the resonant QNMs display two peaks as a function of the axial deformation. The most pronounced one corresponds to the hybridization of the resonant QNMs giving rise to a quasi-BIC mode, while the second is due to the hybridization of a QNM with the background ones. Superscattering in the ED channel (a super dipole mode) arises in the red-shaded regions. Interestingly, the single channel superscattering regime can be achieved in both resonant QNMs at relatively low $\mathcal{Q}$-factors (points A and C), but can also be tuned to a high-$\mathcal{Q}$ near the quasi-BIC (point B), where the dipole strength becomes quenched and the radiation leaks primarily through the MQ channel, (blue-shaded region). In points A and C, the most appreciable signature of ED radiation for the QNMs involved is the appearance of a central hotspot of the electric field inside the nanorod, as well as side lobes. As shown in Fig. \ref{fig_1} and the inset of Fig. \ref{fig_3}(a), at the quasi-BIC, where only MQ radiation is allowed, the central hotspot is absent. In all cases, the incident plane wave is significantly distorted by the scattered field [Fig. \ref{fig_3}(b)]. In the single channel superscattering regime, the ED is shown once again to almost double its established bound (refer to Sec. S5 of the Supplemental Information \cite{Suppl_Mat}). 

{\em Experimental demonstrations.} We perform a proof-of-concept experiment by measuring the extinction cross-section and scattering patterns of disk-shaped particles in the microwave frequency range. We reproduce the geometrical parameters of the rod in Fig.~\ref{fig_3} using a set of ceramic resonators with fixed $4.0$~mm radii, and permittivity $\varepsilon=22$ with loss tangent $0.001$. As shown in the inset of Fig. \ref{fig_5}(a), three samples are assembled from several disks to obtain the desired aspect ratios $\Delta h/r$ for the resonators. 
The results of measurements of both the total extinction cross-section and electric near-field patterns are collected in Fig.~\ref{fig_5} (see details in Sec. S6 of the Supplemental Material \cite{Suppl_Mat}). They are in reasonable agreement with the simulations performed in COMSOL Multiphysics. Since the resonances red-shift with increasing size, the observations were performed in a broader frequency range in Fig. \ref{fig_5}(c). In the highlighted frequencies of Figs.~\ref{fig_5}(a,c), we observe broad resonances with large extinction values, characteristic of the proposed superscattering modes. Indeed, the plane wave is seen to be strongly distorted in the near field [lower panels of Figs.~\ref{fig_5}(a,c)]. The quasi-BIC  appears at the expected value of $\Delta h/r= 0.625$, manifesting itself as a sharp peak in the spectra [Fig. \ref{fig_5}(b)]. The results provide experimental evidence of the control of both the $\mathcal{Q}$-factor and scattered power between two resonances to achieve the superscattering regime.

{\em Conclusions.} We have demonstrated how two interfering resonances in the quasi-BIC regime can be exploited to achieve unexplored superscattering regimes with subwavelength dielectric nonspherical resonators. We have observed the single channel superscattering originating from an electric `super'- dipole moment being two times stronger than the currently established limit. The avoided resonance crossing leads to energy exchange between the radiation channels allowing to control both $\mathcal{Q}$-factors and multipolar contents of resonances while maintaining a high scattering cross-section. Besides their fundamental interest, such exotic scattering can be useful in biosensing \cite{barhom2019biological,kostina2019optical}, energy harvesting \cite{terekhov2019broadband,terekhov2019enhanced,kozlov2016asymmetric}, and can also potentially arise in acoustics.

\begin{acknowledgments}
A.C. and H.K.S. contributed equally to this work. A.S.K. and V.R.T. acknowledge financial support from the National Key R\&D Program of China (Project No. 2018YFE0119900).
\end{acknowledgments}

\bibliography{ms}

\end{document}


\preprint{}

\title{SUPPLEMENTAL MATERIAL\\Superscattering Empowered by Bound States in the Continuum}
\author{Adri\`a Can\'os Valero,$^1$ Hadi~K.~Shamkhi,$^1$ Anton S. Kupriianov,$^2$ Vladimir R. Tuz,$^3$ Alexander A. Pavlov,$^4$ Dmitrii Redka,$^5$ Vjaceslavs Bobrovs,$^6$ Yuri S. Kivshar,$^7$ and Alexander~S.~Shalin$^{1,4,8}$}

\affiliation{$^1$ITMO University, St.~Petersburg 197101, Russia}
\affiliation{$^2$College of Physics, Jilin University, Changchun 130012, China} 
\affiliation{$^3$State Key Laboratory of Integrated Optoelectronics, College of Electronic Science and Engineering, International Center of Future Science, Jilin University, Changchun 130012, China}
\affiliation{$^4$Institute of Nanotechnology of Microelectronics of the Russian Academy of Sciences, Moscow 119991, Russia}
\affiliation{$^5$Electrotechnical University LETI, St. Petersburg 197376, Russia}
\affiliation{$^6$Riga Technical University, Institute of Telecommunications, Riga 1048, Latvia}
\affiliation{$^7$Nonlinear Physics Centre, Australian National University, Canberra ACT 2601, Australia}
\affiliation{$^8$Kotelnikov Institute of Radio Engineering and Electronics of Russian Academy of Sciences (Ulyanovsk branch), Ulyanovsk 432000, Russia}

\maketitle

\tableofcontents

\renewcommand{\thesection}{S\arabic{section}} 
\renewcommand{\theequation}{S.\arabic{equation}}

\section{Derivation of the TCMT model} \label{Sup.A}

Here, we first introduce the general set of equations defining the TCMT, and later proceed with a series of simplifications to recover Eqs. (1), (3), and (4) of the main text. The coupled mode equations can be written as \cite{haus1991coupled,suh2004temporal,fan2003temporal}:
\begin{align}
    \frac{d\ket{\mathbf{\Psi}}}{dt}=-i\mathcal{H}_0\ket{\mathbf{\Psi}}+A^T\mathbf{s}^+,\\
    \mathbf{s}^-=\mathcal{B}\mathbf{s}^++A\ket{\mathbf{\Psi}}.
\end{align}
The Hamiltonian $\mathcal{H}_0$ can be split into Hermitian ($\Omega$) and anti-Hermitian ($-i\Gamma$) parts given by
\begin{align}
    \Omega=\left(\begin{array}{cc}
         \omega_1&\kappa  \\
         \kappa&\omega_2 
    \end{array}\right),\\
    \Gamma=\left(\begin{array}{cc}
         \gamma_1&\nu  \\
         \nu&\gamma_2 
    \end{array}\right).
\end{align}
The matrix $\Omega$ contains the resonant frequencies of a closed cavity with mutually coupled resonances.The matrix $\Gamma$ accounts for radiative energy leakage from the cavity to the continuum, as well as additional radiative coupling between the modes. Arguments based on time reversal symmetry and energy conservation constraints \cite{haus1991coupled} lead to the following relations:
\begin{align}
            A^*=-A \label{eq:sup1},\\
        A^{\dagger}A=2\Gamma \label{eq:sup2}.
\end{align}
In the previous, we have assumed negligible contributions from the non-resonant scattering background by setting $\mathcal{B}=I$ (identity matrix). We make the change of variables $A=i\sqrt{2}D$, where all elements of $D$ are real due to Eq. (\ref{eq:sup1}):
\begin{equation} \label{eq:sup3}
    D=\left ( \begin{array}{cc}
       d_1 &d_2  \\
         d_{01}&d_{02} 
    \end{array}\right ).
\end{equation}
Substituting in Eq. (\ref{eq:sup2}) leads to a relation between $D$ and $\Gamma$:
\begin{equation} \label{eq:sup4}
    D^{T}D=\Gamma,
\end{equation}
so that
\begin{align}
\label{eq:sup5}
    \gamma_{i}=d_{i}^2+d_{0i}^2,\\
    \nu=d_1d_2+d_{01}d_{02}. \label{eq:sup6}
\end{align}
The eigenfrequencies of $\mathcal{H}_0$ are given by
\begin{align} \label{eq:18}
    \tilde{\omega}_u=\frac{1}{2}\left(\Delta\tilde{\omega}_++\sqrt{4g^2+\Delta\tilde{\omega}_{-}^2}\right),\\
    \label{eq:19}
    \tilde{\omega}_d=\frac{1}{2}\left(\Delta\tilde{\omega}_+-\sqrt{4g^2+\Delta\tilde{\omega}_{-}^2}\right).
\end{align}
 In Eqs. (\ref{eq:18}) and (\ref{eq:19}), $\Delta\tilde{\omega}_{\pm}=(\omega_2\pm\omega_1)-i(\gamma_2\pm\gamma_1)$. As discussed in the main text, we consider the following simplifications:
 \begin{align}
    \gamma_1=d_1^2,\\
    \gamma_2=d_1^2+d_0^2,\\
    \nu=\gamma_1.
\end{align}
 Then, a Taylor series over $\kappa$ yields, to first order, more treatable expressions for Eqs. (\ref{eq:18}) and (\ref{eq:19}):
\begin{align}
    \label{eq:30}
    \Re(\tilde{\omega}_{u,d})=\omega_1\pm{}\frac{2\kappa}{\sqrt{4+\alpha^4}}+O(\kappa^2),\\
    \label{eq:31}
    \Im(\tilde{\omega}_{u,d})=-d_1^2\Big[(2+\alpha^2)\mp{}\sqrt{4+\alpha^4}\Big]+O(\kappa^2),
\end{align}
where $\alpha=d_0/d_1$.

The $\mathcal{T}$-matrix of the two-level model in the harmonic regime can be obtained in a simple fashion:
\begin{equation} \label{Eq:sup7}
    \mathcal{T}=iD(\mathcal{H}_0-I\omega)^{-1}D^{T}.
\end{equation}
The $\mathcal{T}_{11}$ and $\mathcal{T}_{12}$ components take the form:
\begin{align}
    \mathcal{T}_{11}(\omega)=i\frac{d_2^2(\tilde{\omega}_1-\omega)+d_1^2(\tilde{\omega}_2-\omega)-2d_1d_2g}{(\tilde{\omega}_1-\omega)(\tilde{\omega}_2-\omega)-g^2},\\
    \mathcal{T}_{12}(\omega)=i\frac{d_0d_2(\tilde{\omega}_1-\omega)-d_1d_0g}{(\tilde{\omega}_1-\omega)(\tilde{\omega}_2-\omega)-g^2}.
\end{align}
We consider the modulus squared of each component:
\begin{align}
    |\mathcal{T}_{11}(\omega)|^2=d_1^4\frac{(\gamma_1-\gamma_2)^2+4(\omega_1-\kappa-\omega)^2}{|(\tilde{\omega}_1-\omega)(\tilde{\omega}_2-\omega)-g^2|^2},\\
     |\mathcal{T}_{12}(\omega)|^2=d_1^2d_0^2\frac{(\omega_1-\kappa-\omega)^2}{|(\tilde{\omega}_1-\omega)(\tilde{\omega}_2-\omega)-g^2|^2},
\end{align}
with
\begin{equation}
      \norm{(\tilde{\omega}_1-\omega)(\tilde{\omega}_2-\omega)-g^2}^2=[\gamma_1(\gamma_1-\gamma_2)-\kappa^2+(\omega-\omega_1)^2]^2+[(\gamma_1+\gamma_2)(\omega-\omega_1)+2\gamma_1\kappa]^2.
\end{equation}
We can evaluate the $\mathcal{T}$-matrix components at $\omega_1$ to get the total contribution to the first scattering channel
\begin{align}
\label{eq:32}
    \frac{\sigma_1}{\sigma_0}=\frac{d_1^4\left[d_0^4+\kappa^2(2+\alpha)^2\right]}{(d_1d_0)^4+2\kappa^2(2+\alpha^2)+\kappa^4},
\end{align}
where we have assumed for simplicity that the two input coefficients are identical. A Taylor expansion of Eq. (\ref{eq:32}) with respect to $\alpha$ yields
\begin{equation}
    \frac{\sigma_1}{\sigma_0}=\frac{4\gamma_1^2(1+\alpha)}{4\gamma_1^2+\kappa^2}+O(\alpha^2).
\end{equation}
We note two important assumptions in Eq. (\ref{eq:32}); first, the $D$ matrix was derived considering the idealized situation where no background is present. The scattering background in an isolated resonator is produced by the influence of QNMs lying outside the spectral range of interest, as well as zero-frequency modes. The $d_i$ coefficients would become complex. However, this merely complicates the analytical expressions, since energy conservation is fulfilled in all cases, and therefore the conclusions regarding the scattering limit are unaffected. Furthermore, the CMT equations in this work are used as a theoretical guideline and not as a design tool, since exact cavity PT and direct numerical calculations of the $\mathcal{T}$-matrix provide a more quantitative framework for our findings. Second, the two input channels were assumed to have the same weight. In the case of an incident plane wave, the input channels correspond to vector spherical harmonics, and therefore their weights will be different, and their ratio can be possibly complex (refer to Section S2).

\section{Normalization of multipolar channels for TCMT} 

We start with the expansion for an arbitrary field \cite{jackson1999classical}:
\begin{align}
    \label{eq:5}
    \mathbf{E}(\mathbf{r})=\sum_{lm}\left[a_{lm}\mathbf{N}_{lm}+b_{lm}\mathbf{M}_{lm}\right],\\
    Z_0\mathbf{H}(\mathbf{r})=\sum_{lm}\left[-a_{lm}\mathbf{M}_{lm}+b_{lm}\mathbf{N}_{lm}\right],
    \label{eq:6}
\end{align}
where the coefficients are in units of electric field. For the TCMT to be fully consistent, we require them to be in units of $\sqrt{\text{power}}$. Thus, we can rewrite Eq. (\ref{eq:5}) in terms of the new coefficients:
\begin{equation}
    \label{eq:7}
    \mathbf{E}(\mathbf{r})=A_0\sum_{lm}\left[a^p_{lm}\mathbf{N}_{lm}+b^p_{lm}\mathbf{M}_{lm}\right],
\end{equation}
where $A_0=k\sqrt{2Z_0}$. In this fashion, $|a_{lm}^p|^2$ is in units of power, and the same for the $b_{lm}^p$. 

The VSH expansion of a normally incident plane wave is given by the normalized coefficients:
\begin{align}
    \label{eq:8}
    b^p_{\pm,lm}=\frac{i^{l+1}k}{\sqrt{Z_0}}\sqrt{2\pi(2l+1)}\delta_{m,\pm1},\\
    \label{eq:9}
    a^p_{\pm,lm}=\pm{}b^p_{\pm,lm}.
\end{align}
From here one can directly infer the single channel limitation if no mixing takes place within the scatterer, since it is determined by the multipolar content of the plane wave:
\begin{equation} \label{eq:10}
     |a^p_{\pm,lm}|^2=\frac{\lambda^2(2l+1)}{\pi}.
\end{equation}
We note that Eq. (\ref{eq:10}) corresponds to the limitation under circularly polarized plane wave illumination. A linearly polarized plane wave can be straightforwardly deduced from the linear combination of the two helicities, i.e. $\mathbf{e}_x=(\mathbf{e}_++\mathbf{e}_{-})/2$,
\begin{equation} \label{eq:11}
    \frac{1}{4}(|a^p_{+,lm}|^2+|a^p_{-,lm}|^2)=\frac{\lambda^2(2l+1)}{2\pi}.
\end{equation}
The normalized multipole coefficients of a linearly polarized plane wave are:
\begin{align}
    \label{eq:12}
    a^p_{lm}=\frac{i^{l+1}k}{2\sqrt{Z_0}}\sqrt{2\pi(2l+1)}(\delta_{m,+1}-\delta_{m,-1}),\\
    b^p_{lm}=\frac{i^{l+1}k}{2\sqrt{Z_0}}\sqrt{2\pi(2l+1)}(\delta_{m,+1}+\delta_{m,-1}).
    \label{eq:13}
\end{align}
The electric dipole is:
\begin{equation} \label{eq:31}
    a^p_{1m}=-k\sqrt{\frac{3\pi}{2Z_0}}(\delta_{m,+1}-\delta_{m,-1}),
\end{equation}
and the magnetic quadrupole is:
\begin{equation} \label{eq:32}
    b^p_{2m}=-ik\sqrt{\frac{5\pi}{2Z_0}}(\delta_{m,+1}+\delta_{m,-1}).
\end{equation}
For a given $m$, the ratio between the two partial amplitudes is simply $b^p_{2\pm}/a^p_{1\pm}=\pm{}i\sqrt{5/3}$. The relation between the plane wave coefficients  and the scattered multipoles is directly given by the $\mathcal{T}\text{-matrix}$. Considering only the electric dipole and magnetic quadrupole, normalizing all amplitudes by $a^p_{1\pm}$ yields, finally:
\begin{equation} \label{eq:16}
   \mathbf{p}=\mathcal{T} \left(\begin{array}{c}
         1  \\
         \pm{}i\sqrt{5/3} 
    \end{array}\right).
\end{equation}

\section{Maximizing single channel superscattering} 

\renewcommand{\thefigure}{S\arabic{figure}}
\begin{figure*}[!b]
    \centering
    \includegraphics[scale=0.8]{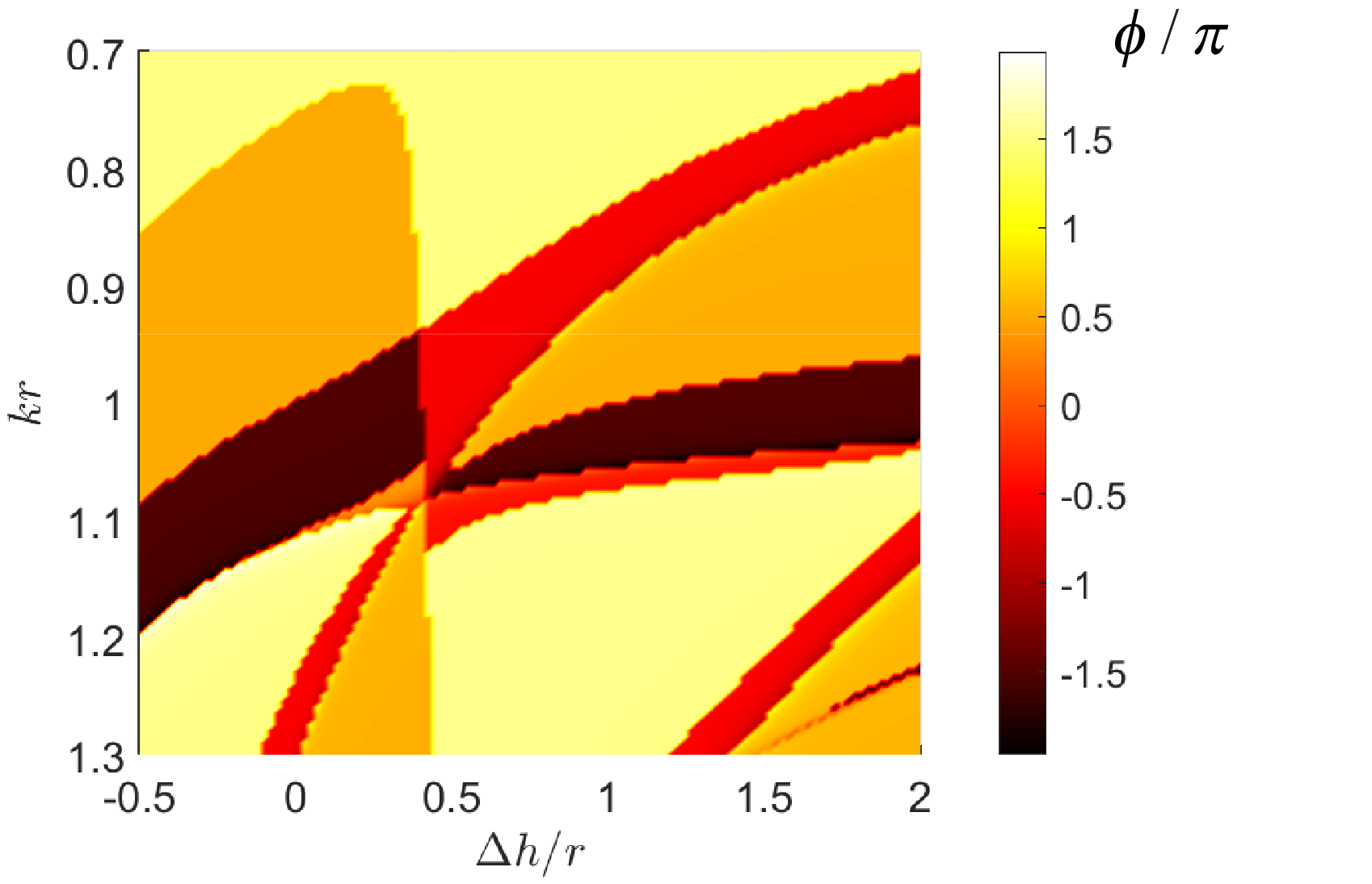}
    \caption{Phase difference between $\mathcal{T}_{11}$ and $\mathcal{T}_{12}$. The azimuthal number in this case is $m=1$.}
    \label{fig S1}
\end{figure*}

In this section we prove a simple condition that leads to overcoming the single channel limit when the $\mathcal{T}$-matrix is non-diagonal. While here we treat the case of two coupled channels in a cylindrically symmetric scatterer, similar conditions can be obtained for an arbitrary number of them. This lies outside the scope of the work. 
First, energy conservation leads to a well-known relation for the components of the $\mathcal{T}$-matrix:
\begin{equation} \label{eq:34}
    2\mathcal{T}^{\dagger}\mathcal{T}=\mathcal{T}^{\dagger}+\mathcal{T}.
\end{equation}
This relation limits the maximum attainable values of the components in $\mathcal{T}$. E.g., $\norm{\mathcal{T}_{ii}}\leq{}1$ when $\mathcal{T}_{ii}$ is real. However, the total scattering from one channel is given by
\begin{equation}
    \norm{p_i}^2=\norm{a_i\mathcal{T}_{ii}+a_j\mathcal{T}_{ij}}^2,
\end{equation}
which can be rewritten as
\begin{equation}
    \norm{p_i}^2=\norm{a_i}^2\norm{\mathcal{T}_{ii}}^2+\norm{a_j}^2\norm{\mathcal{T}_{ij}}^2+2\norm{\mathcal{T}_{ii}}\norm{\mathcal{T}_{ij}}[\Re(a_ia_j^*)\cos(\theta_{ii}-\theta_{ij})-\Im(a_ia_j^*)\sin(\theta_{ii}-\theta_{ij})].
\end{equation}
The first two terms are always positive, and can be maximized in the vicinity of a resonance, but the other two terms can also be negative depending on the $a_i$ and the phase difference $\theta_{ii}-\theta_{ij}$. It can be physically interpreted as the interference between the partial contributions to the channel $i$. Maximizing single channel superscattering thus reduces to maximizing $\phi^{+}(i,j)=\Re(a_ia_j^*)\cos(\theta_{ii}-\theta_{ij})$ or $\phi^{-}(i,j)=-\Im(a_ia_j^*)\sin(\theta_{ii}-\theta_{ij})$ . With the help of the energy-normalized plane wave coefficients given in Eqs. (\ref{eq:12}), and (\ref{eq:13}), for $m=1$ these two functions can be shown to be proportional to
\begin{align} \label{eq:37}
    \phi^+(i,j)\propto{}\cos(\pi/2(\ell_{i}-\ell_{j}))\cos(\theta_{ii}-\theta_{ij}),\\\label{eq:38}
    \phi^-(i,j)\propto{}-\sin(\pi/2(\ell_{i}-\ell_{j}))\sin(\theta_{ii}-\theta_{ij}).
\end{align}

If the $\mathcal{T}$-matrix couples two multipoles with an even (odd) difference of angular momentum per photon $\Delta\ell_{ij}=\ell_i-\ell_j$, $\phi^{\pm}(i,j)=0$. Thus, to maximize single channel superscattering for even $\Delta\ell_{ij}$, $\phi^{+}(i,j)$ must be maximized, and vice versa for odd $\Delta\ell_{ij}$. We note that $\phi^{\pm}(i,j)$ is limited to 1. Depending on the sign of the first term multiplying the rhs of Eqs. (\ref{eq:37}) and (\ref{eq:38}), the maximum single channel superscattering will occur when the phase difference between the $\mathcal{T}$-matrix components is either 0 or $\pi$ for even $\Delta\ell_{ij}$, or $\pm\pi/2$ for odd. 

To validate the previous considerations, we calculate the phase difference between the $\mathcal{T}$-matrix components of the nanoresonator studied in the main text. We focus on the energy exchange between the electric dipole (scattering channel 1, $\ell=1$) and magnetic quadrupole (scattering channel 2, $\ell=2$). According to the rules established in Eqs. (\ref{eq:37}) and (\ref{eq:38}), a super dipole will occur when the phase difference between the off-diagonal component and the direct dipole contribution is either $+$ or $-\pi/2$. The calculations indeed demonstrate that this is the case (Fig. \ref{fig S1}), since a phase difference of $-\pi/2$ is observed in the regions where the dipole scattering is maximized (refer to Fig. 4(a) in the main text). In the example, the azimuthal number is $m=1$. We remark that we have implicitly assumed that the scatterer has cylindrical symmetry, and therefore only multipoles with equal $m$ can couple.
\section{Multipolar decomposition and eigenmodal content of a nanosphere}
\begin{figure*}[!htbp]
    \centering
    \includegraphics[scale=0.7]{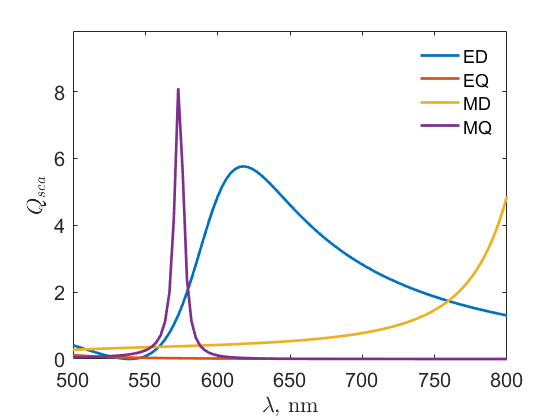}
    \caption{Multipolar decomposition of the scattering efficiency up to the quadrupoles for a silicon nanosphere ($n\approx 4$) with radius $a=$100 nm in the visible range.}
    \label{fig_S2}
\end{figure*}

The multipolar response of a sphere can be recovered analytically with the help of Mie theory. Specifically, the electric and magnetic coefficients take the well-known form \cite{bohren2008absorption}:
\begin{align}
    \label{eq:39}
    a_{\ell}=\frac{m \psi_{\ell}(m x) \psi_{\ell}^{\prime}(x)-\psi_{\ell}(x) \psi_{\ell}^{\prime}(m x)}{m \psi_{\ell}(m x) \xi_{\ell}^{\prime}(x)-\xi_{\ell}(x) \psi_{\ell}^{\prime}(m x)},\\ \label{eq:40}
    b_{\ell}=\frac{\psi_{\ell}(m x) \psi_{\ell}^{\prime}(x)-m \psi_{\ell}(x) \psi_{\ell}^{\prime}(m x)}{\psi_{\ell}(m x) \xi_{\ell}^{\prime}(x)-m \xi_{\ell}(x) \psi_{\ell}^{\prime}(m x)},
\end{align}
where $m$ is the ratio between the refractive index of the sphere and the outer medium, and $x=ka$, $a$ being the radius of the sphere. The $\psi_{\ell}(z)$, $\xi_{\ell}(z)$ are Ricatti-Bessel functions \cite{bohren2008absorption}. For a sphere, the QNMs are classified in electric and magnetic, with different $\ell$ \cite{fuchs1968optical}. The eigenfrequencies correspond to the poles of Eqs. (\ref{eq:39}) and (\ref{eq:40}), which can be obtained by making the denominators zero, giving the characteristic equations:
\begin{align}
    \label{eq:41}
    \frac{\left[x h_{\ell}^{(1)}(x)\right]^{\prime}}{h_{\ell}^{(1)}(x)}=\frac{\left[m x j_{\ell}(m x)\right]^{\prime}}{ m^{2} j_{\ell}(m x)},\\
    \frac{\left[x h_{\ell}^{(1)}(x)\right]^{\prime}}{h_{\ell}^{(1)}(x)}=\frac{\left[m x j_{\ell}(m x)\right]^{\prime}}{j_{\ell}(m x)},
\end{align}
for the electric and magnetic modes, respectively. In the previous, $j_{\ell}(x)$ and $h_{\ell}^{(1)}(x)$ are, respectively, the spherical Bessel and Hankel functions \cite{bohren2008absorption}. For the nanosphere in the main article, only two QNMs are resonant in the visible range, with eigenfrequencies given by $\tilde{x}_1=1.05273-0.07236i$ (ED, $\ell=1$) and $\tilde{x}_2=1.09572-0.006840i$ (MQ, $\ell=2$). The positions of the poles indeed coincide with the peaks of the ED and MQ of the multipole decomposition shown in Fig. \ref{fig_S2}. 

\section{Super-ED modes of a silicon nanorod}

\begin{figure*}[!htbp]
    \centering
    \includegraphics[scale=0.73]{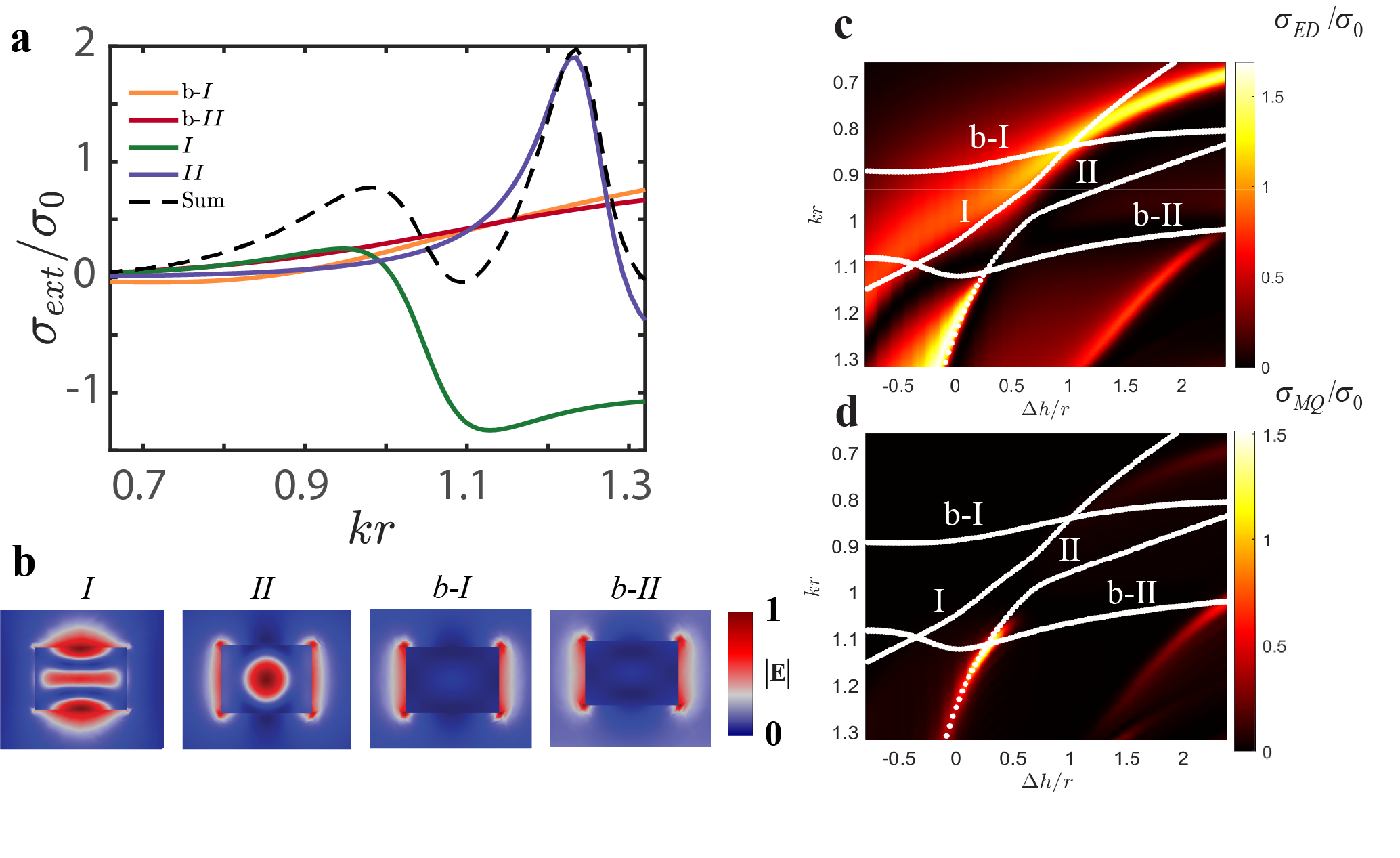}
    \caption{(a) QNM decomposition of the extinction cross section from a silicon nanorod with radius 126 nm and height 180 nm, corresponding to the unperturbed rod whose QNMs are used as a basis for the axial deformation in the main text. The QNMs I, II are resonant, while b-I and b-II constitute to the background. (b) Field distributions of the involved QNMs. (c) Normalized scattering cross section of the ED channel. The superimposed white traces correspond to the paths followed by the resonant frequencies of the involved QNMs, analytically recovered with cavity perturbation theory with the reduced QNM set shown in~(b).}
    \label{fig_S3}
\end{figure*}

In this section we present the results for the QNM analysis of the silicon nanorod discussed in the main text [Figs. \ref{fig_S3}(a) and \ref{fig_S3}(b)], as well as their multipolar response at real frequencies [Figs. \ref{fig_S3}(c) and \ref{fig_S3}(d)]. The calculations are performed by means of COMSOL Multiphysics and the complementary Matlab scripts provided in the package MAN \cite{yan2018rigorous}. As expected, the fields of the resonant modes are strongly concentrated within the particle, while the background modes are mainly surface modes with low $\mathcal{Q}$-factor, resembling plasmonic modes.
Scattering in the ED channel can be enhanced up to 1.6 times its value in a sphere [see Fig. \ref{fig_S3}(c)]. We note that reciprocity does not allow a complete suppression of the MQ channel radiation, but the latter can be rendered up to 3 times smaller than the ED scattering [please refer to Fig. \ref{fig_S3}(d)]. The paths traced by the eigenfrequencies calculated by cavity perturbation theory [superposed in [Figs. \ref{fig_S3}(c) and \ref{fig_S3}(d)] coincide very well with the full numerical simulations, with deviations appearing at large deformations due to having considered only the first order term in the perturbation Hamiltonian. The results show that the chosen QNM basis is sufficient for a quantitative description of the effect.
\section{Experimental section} \label{Sup.5}

A Taizhou Wangling TP-series microwave ceramic composite is used as a dielectric material for resonators (disks) fabrication. According to the manufacturer data sheet, relative permittivity of this ceramic is $22 \pm 1$ and loss tangent is $\tan \delta \approx 1 \cdot 10^{-3}$ at $10$~GHz. A set of disks is fabricated from the ceramic plates with the use of a precise milling machine. The radius of disks is 4 mm. The sample to be measured is composed of several disks to obtain several resonators with thicknesses of 6.0, 7.5, and 10.0 mm. The sample is mounted on a support made of ROHACELL\textsuperscript{\textregistered} 71~HF plate, whose relative permittivity is $1.01\pm0.05$.

\begin{figure*}[!h]
    \centering
    \includegraphics[scale=0.3]{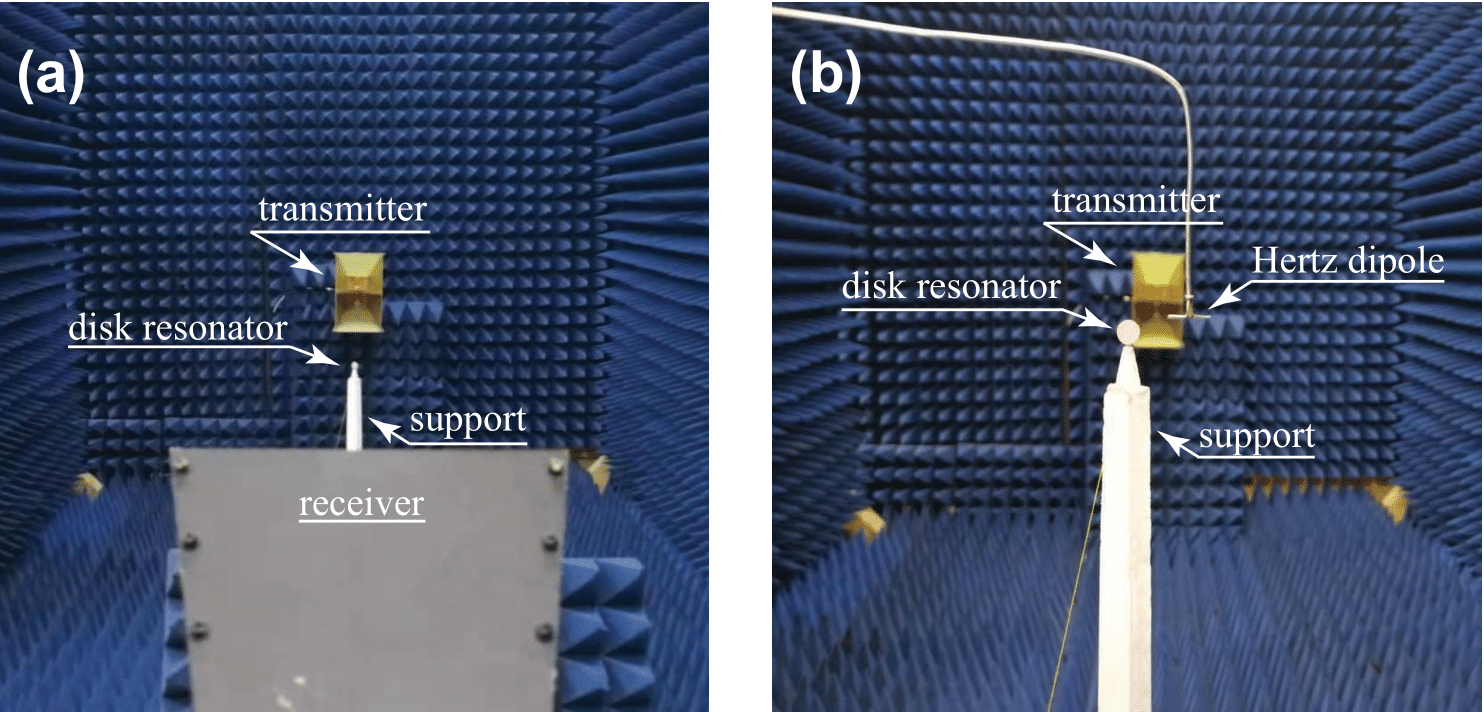}
    \caption{Experimental setups for (a) measurement of the extinction cross-section and (b) mapping of the electric near-field pattern.}
    \label{fig_S4}
\end{figure*}

To measure the total extinction cross-section, the sample is placed in an echoic chamber and illuminated by a normally incident linearly-polarized waves radiated and received by a pair of HengDa Microwave HD-10180DRA10 horn antennas. A photograph of the experimental setup is shown in Fig. \ref{fig_S4}(a). The antennas operating range is 1--18~GHz. The distance between each horn antenna and the sample is fixed at 2.0~m. The antennas are connected to the ports of Rhode \& Schwarz ZVA50 Vector Network Analyzer (VNA) by 50-$\Omega$ coaxial cables. Using VNA the $S21$-parameter (transmission coefficient) is detected and analyzed by a special computer software. Forward scattering is obtained from the measured transmission coefficient. The total extinction cross-section is extracted from the measured complex magnitude of the forward-scattered signal by means of the optical theorem \cite{Newton_AmJPhys_1976}. A stainless steel sphere with the radius $40.00 \pm 0.04$ mm and the corresponding analytical Mie solution are used for the signal calibration. To reduce unwanted reverberations between antennas the time gating technique is used. Then a standard filtering procedure is applied to the raw data to remove the background noise. 

A photograph of the measurement setup for mapping the electric near-field is presented in Fig. \ref{fig_S4}(b). The transmitting antenna generates a linearly polarized quasi-plane-wave whose polarization is orthogonal to the symmetry axis (the $z$ axis) of the disk resonator. The receiving antenna is substituted by an electrically small probe. This probe is a Hertz dipole which is oriented along the polarization direction of the incident wave. The probe detects the dominant component ($E_x$) of the scattered electric near-field. A LINBOU near-field imaging system is used for the near-field mapping. The scan area around the sample has dimensions $50 \times 50$ mm$^2$ in the $x$-$z$ plane. In the measurements, the probe automatically moves in the scan area with a 1-mm step along two orthogonal directions. At each probe position, both the amplitude and the phase of the scattered electric field are measured. The exception is the area near the sample. This area is experimentally inaccessible because of the technical limitation on the distance between the moving probe and the vertically standing sample.

\bibliography{supplement}